 \documentclass[aps,ams,12pt,onecolumn,preprint,nofootinbib,showpacs,superscriptaddress]{revtex4}

\usepackage{graphicx}\usepackage{epsfig}
\usepackage[psamsfonts]{amssymb}
\usepackage{amsfonts}\usepackage{amscd}
\usepackage{amsmath,amssymb}
\usepackage{cases}

\newcommand{\ba}{\begin{eqnarray}}
\newcommand{\ea}{\end{eqnarray}}

\def \Imm { \mbox{\rm Im} }
\def \Ree { \mbox{\rm Re} }

\everymath{\displaystyle}
\graphicspath{{./Figures-2+2/}}

\begin{document}
\thispagestyle{myheadings}


\title{ Analytical calculations of the  tenth order QED radiative corrections to  lepton anomalies
within the Mellin--Barnes representation}

\author{O.P. Solovtsova}
\email{olsol@theor.jinr.ru; solovtsova@gstu.gomel.by}
\affiliation{Bogoliubov Lab. Theor. Phys., JINR, Dubna, 141980,
Russia}
 \affiliation{Gomel State Technical
University, Gomel, 246746, Belarus}
\author{V.I. Lashkevich}
\affiliation{Gomel State
Technical University, Gomel, 246746, Belarus}
\author{L.P. Kaptari}
\email{kaptari@theor.jinr.ru} \affiliation{Bogoliubov Lab.
Theor. Phys., JINR, Dubna,
141980, Russia}

\begin{abstract}

We investigate the radiative quantum electrodynamic (QED) corrections to the lepton ($L=e, ~\mu $ and $\tau$)  anomalous magnetic moment due to
the contributions of diagrams with insertions of the photon
vacuum polarisation operator  consisting solely of  four closed lepton  ($l=e, ~\mu $ and $\tau$) loops. Moreover, we focus on specific  operators with  two loops   formed by  leptons $L$ of the same type as the external one, the other two formed by leptons $\ell$ different from $L$.
The approach is essentially based on the employment of the Mellin--Barnes representation of the $x$-parametrization
of the corresponding Feynman diagrams. This allows one to obtain, for the first time,  exact analytical expressions for the radiative corrections of the tenth order w.r.t. the electromagnetic coupling constant $e$.
Analytically, the radiative corrections are expressed in terms of the ratio $r=m_\ell/m_L$ of the internal to external lepton masses.
The dependence  on $r$ is investigated numerically in  the whole interval of $r$, $0 < r < \infty$. To make comparisons with earlier published results possible, our exact analytical expressions are expanded about $r=0$ and $r \to \infty$ and compared with the corresponding asymptotic  expansions known in the literature.

\end{abstract}



\maketitle

\section{Introduction}

Among the most important consequences of the Dirac theory is the prediction~\cite{dirac} that the gyromagnetic factor $g_L$ of a lepton  $L$ is $g_L=2$. However, the   self-interaction with photons leads to a gyromagnetic factor $g_L\neq 2$, which in the literature is referred to as the lepton anomaly, $a_L=(g_L-2)/2\neq 0$. Obviously, this anomaly is an important characteristic of the magnetic field surrounding a lepton and, in spite of its extremely small deviation from zero, it can serve as a substantial test of the Standard Model (SM) or even  can  indicate the  existence of some ``new physics'' beyond the SM. Clearly, the self-energy correction to the lepton electromagnetic vertex originates not only from the electromagnetic interaction but also  from strong and weak interactions. A comprehensive review of contributions of different mechanisms to $a_L$ can be found in, e.g., Refs.~\cite{Jegerlehner:2017gek,review-2021}. At present, experimental measurements of $a_L$  for electrons~\cite{Parker:2018vye,Morel} and muons~\cite{E989,Fermilab2023} are performed with an extremely high  accuracy  which imposes appropriate  requirements on  theoretical calculations. First theoretical calculations of the leading order corrections were performed by J.~S.~Schwinger~\cite{Schwinger1948} and found to be $a_L=\alpha/2\pi$, where $\alpha$ is the fine structure constant. Next to the leading order corrections involve much more diagrams which result in complicate and cumbersome calculations~\cite{Petermann:1957hs,Sommerfield:1957zz}. Usually, high precision numerical calculations of the radiative corrections are performed numerically by expressing a full set of Feynman diagrams via a number of  master integrals associated  to all possible, at a given order, topologies with a subsequent analytical fit by  the known PSLQ algorithm~\cite{PSLQ1992}.
So the full set of the eighth order corrections contains more than 890 diagrams, which can be  expressed by means of 334 master integrals belonging to 220 topologies. Then applying the PSLQ algorithm to the  master integrals, one achieves a very high precision (more than 1100 digits) for the corresponding corrections, cf.~Refs.~\cite{Laporta:2017okg,elliptic1}. However, such numerical calculations are rather computer resources consuming and, in addition, detailed investigations of the role of different mechanisms contributing to $a_L$ are hindered. Therefore, it is enticing to find at least   a subset of  specific  Feynman diagrams which can provide analytical expressions even if only for a restricted number of perturbative  terms. Then, having at hand analytical expressions, one can perform calculations with any desired  accuracy and, consequently,  use  as excellent tests of the reliability of direct numerical procedures. In addition, a detailed analysis of the contribution of different lepton loops to $a_L$ will become a possible. It turns out that the subset of diagrams with  loops originating only from  insertions of the photon polarisation operator, the so-called ``bubble''-like diagrams, allows for analytical calculations of corrections up to fairly high orders. The distinctive peculiarity of this subset is that each diagram with an arbitrary number of inserted loops can be expressed via one diagram of the second order with the exchange of only one but  massive photon. The explicit expression for these kinds of massive diagrams is well known in the literature~\cite{berestetski,brodski} and its analytical form permits further mathematical manipulations to obtain final analytical formulae for each considered diagram.
The main idea is to apply the dispersion relations to the corresponding Feynman diagram to express it via the Feynman  $x$-parametrization  of the second-order diagram with massive photons and finally to apply the Mellins--Barnes representation to the massive photon propagator and again the dispersion relations to the polarisation operators  of the internal  lepton $\ell\neq L$ different from the external one. In this way, one can express any diagram from the mentioned subset in a rather simple form as a comvolution integral of two Mellin momenta (for details, cf. Ref.~\cite{Solovtsova23}). Then, if possible, this integral is carried out analytically providing  the desired exact expressions for $a_L$.

As the first formulations of the approach, one can mention Ref.~\cite{Aguilar:2008qj}, where the corrections to the muon anomaly of the eighth and tenth order were calculated in analytical form as asymptotic expansions at $r=m_\ell/m_L\ll 1$ and $r\to \infty$. Further generalisation of the method for the anomalous magnetic moment of any kind of leptons to obtain exact analytical expressions  up to the eighth order for any possible combination of lepton masses, i.e. for arbitrary $r$ ranging in the interval $(0 < r < \infty)$, was reported in detail in Ref.~\cite{Solovtsova23}.

This paper is a necessary continuation  of our previous studies~\cite{Solovtsova23} of the lepton anomaly within the Mellin-Barnes representation  and  is dedicated  to  analytical  investigations of the tenth order corrections from Feynman diagrams with insertions of four lepton loops. Hitherto corrections of the   order $\alpha^5$ have been considered  either in the asymptotic limit $r\ll 1$~\cite{Aguilar:2008qj} or only for  diagrams with four lepton loops, all  of the same type~\cite{vesti}. Below we consider another type of four loop  diagrams, namely, the ones with two leptons $L$ of the same kind as the external one, the other two leptons $\ell$ different from $L$. The general formalism used in this paper is basically the one reported before in Refs.~\cite{Aguilar:2008qj,Solovtsova23}.

Our paper is organised as  follows. In order to facilitate the reading of the paper, in Section~\ref{sec1} we briefly recall the main definitions relevant to calculations of the lepton anomaly. The general formula for  QED corrections to $a_L$ from the Feynman diagrams with the insertion of an arbitrary number $n$ of lepton loops of two kinds is derived. The resulting convolution integral of two Mellin momenta determining the anomaly $a_L$ is presented explicitly in the most general form. The explicit expression  for each Mellin momentum is derived in Section~\ref{glava2}. The details of calculations of the integral by the Cauchy theorem are discussed in  Subsections~\ref{subsec2} and~\ref{subsec3} in the left ($r<1$) and right ($r>1$) semiplanes of the complex Mellin variable $s$, respectively. The results of numerical calculations of corrections $a_L$ for each type of leptons $e$, $\mu$ and $\tau$ with the insertion of loops with all possible combinations of internal leptons ($e$, $\mu$ and $\tau$) together with a brief discussion  are presented in Section~\ref{numerical}. In Section~\ref{asym}, the exact analytical expressions for $a_L$ are expanded in Taylor series at $r\ll 1$ and $r\gg 1$ and compared with the corresponding asymptotic expansions known in the literature and with the numerical result of exact calculations. The conclusions and summary are presented Section~\ref{summ}.

\section{Basic formalism} \label{sec1}
\begin{figure}
\includegraphics[width=0.88\textwidth]{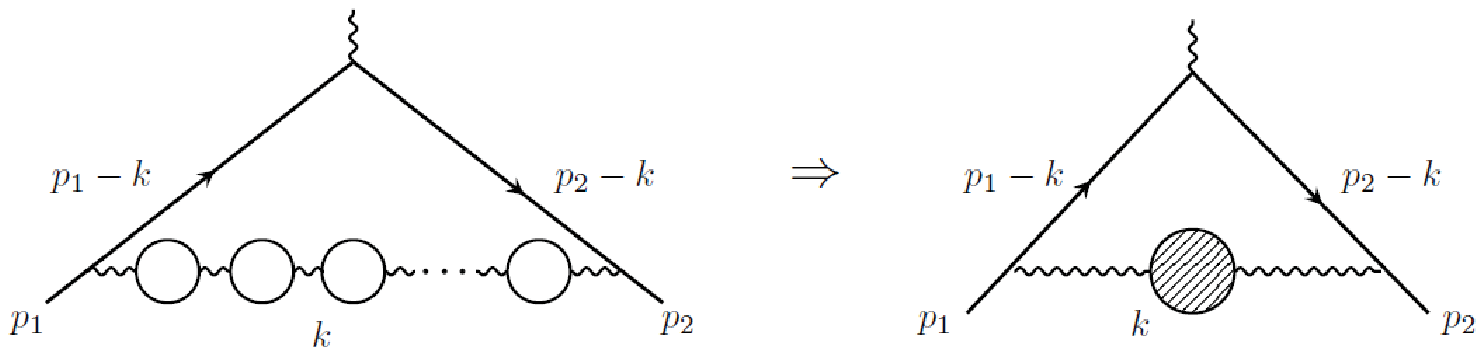}
\vspace*{1mm}
\caption{The Feynman diagrams considered in this paper contributing to the lepton anomalous magnetic moment. Left panel: radiative corrections to the electromagnetic lepton vertex
with insertions of the vacuum polarisation operator with an arbitrary number
$n$ of lepton loops. Right panel:
the second order diagram representing the set of graphs depicted in the left panel as exchanges of one massive
photon, see the text.} \label{Risodin}
\end{figure}

In this Section, we consider the most general form of the QED corrections to the lepton anomalous magnetic moment due to bubble-like Feynman diagrams with the insertion of the photon polarization operator with an arbitrary number $n=p+j$ of loops, where $p$ is the number of loops formed by  leptons $L$ of the same type as the external one, $j$ denotes the leptons $\ell\neq L$. The corresponding Feynman diagram  is depicted in Fig.~\ref{Risodin}, left panel. It is straightforward  to show~\cite{Lautrup:1969fr} (see also \cite{Aguilar:2008qj,Solovtsova23})
that  the electromagnetic vertex can be related to the vertex diagram of the second order with exchanges of one but massive photon, right panel in Fig.~\ref{Risodin}. Direct calculation of the electromagnetic vertex $ \Gamma_\mu(p_1,p_2)$ on the left diagram in Fig.~\ref{Risodin} provides
 \ba
&& \Gamma_\mu(p_1,p_2) = -ie\frac{e^2}{(2\pi)^4}\int d^4k
\gamma_\alpha\frac{(\hat {p_2} -\hat{k} +m_L)\gamma_\mu (\hat {p_1}
-\hat{k} +m_L)} {(k^2-2p_2k)(k^2-2p_1k)}\gamma_\alpha\frac{
\widetilde\Pi(k^2) }{k^2} \\ [0.3cm]&&
 = -ie\frac{e^2}{(2\pi)^4}\int \frac{dt}{t}
\frac1\pi\frac{\Imm \widetilde\Pi(t)}{k^2-t}\int d^4k
\gamma_\alpha\frac{(\hat {p_2} -\hat{k} +m_L)\gamma_\mu (\hat {p_1}
-\hat{k} +m_L)}
{(k^2-2p_2k)(k^2-2p_1k)}=\nonumber\\ [0.3cm]
&&
\frac{1}{\pi} \int \frac{dt}{t}
 {\Imm} \widetilde\Pi(t)\underbrace{\left[ -ie\frac{e^2}{(2\pi)^4}\int d^4k
\gamma_\alpha\frac{(\hat {p_2} -\hat{k} +m_L)\gamma_\mu (\hat {p_1}
-\hat{k} +m_L)} {(k^2-2p_2k)(k^2-2p_1k)}\gamma_\alpha
\frac{1}{k^2-t}\right]}_{\Gamma_\mu^{(2)}(p_1,p_2,t)}
\label{massive} \ea
where $\Gamma_\mu^{(2)}(p_1,p_2,t)$ is the lepton electromagnetic vertex for the second order diagrams with the exchange of one massive photon with  mass  $m_\gamma^2=t$, as depicted in  Fig.~\ref{Risodin}, right panel.

Consequently, the anomalous magnetic moment $a_L$ is directly related to the corresponding anomalous magnetic moment induced by such a massive photon. The latter is well known in the~literature~\cite{berestetski,brodski}:
\vspace{0.2cm}
  \ba a_L&&=\frac1\pi\int \frac{dt}{t}
 \Imm \widetilde\Pi(t) a_L(t)
 =\frac1\pi\int \frac{dt}{t}
 \Imm \widetilde\Pi(t)\frac\alpha\pi\int dx\frac{x^2 (1-x)}{x^2+(1-x)t/m_L^2}=
 \nonumber \\[0.3cm]
  && -\frac\alpha\pi\int dx (1-x)
 \widetilde\Pi\left(-\frac{x^2}{1-x}m_L^2\right)
 .\label{double}
\ea
The last expression  was obtained by exchanging the integration   variables
$dt$ and $dx$ and employing the ``inverse'' dispersion relations to $\Imm\widetilde\Pi(t)$.
Taking into account that the full polarisation operator $\widetilde\Pi(k^2)$ can be written as
\ba \widetilde\Pi(k^2)
=\Pi(k^2)-\Pi^2(k^2)+\Pi^3(k^2)-\cdots \,  \label{power}\ea
and that each operator $\Pi(k^2)$ is a sum of operators of leptons $L$ and $\ell$, the corrections of the $2(n+1)$ order can be expressed in the form of products of the polarisation operators $L$ and $\ell$ each  to the corresponding power. Recall that in our notation, the number of loops $n$ is $n=p+j$; consequently, the polarisation operator $L$  acquires the power $p$ while the  polarisation operator $\ell$ acquires the power $j$. The last effort is to apply the dispersion relations to the $j$-th power of the polarisation operator of the lepton $\ell$ and  then the Mellin--Barnes representation~\cite{Mellin1,Mellin2,Mellin3}. Then one arrives at the final formula~\cite{Aguilar:2008qj,Solovtsova23}
 \ba &&
a_L(p,j)=\frac{\alpha}{\pi} \frac{1}{2\pi i}F_{(p,j)}
\int\limits_{c-i\infty}^{c+i\infty} ds \;
 \left( \frac{4m_{\ell}^2}{m_L^2}\right)^{-s} \Gamma(s)\Gamma(1-s)\;
\left(\frac{\alpha}{\pi}\right)^{p}\Omega_p(s)
\left(\frac{\alpha}{\pi}\right)^{j}R_j(s), \label{fin1}
\ea
where the factor $F_{(p,j)}$ is related to the binomial coefficients $C_{p+j}^p$ as
$F_{(p,j)} =(-1)^{p+j+1} C_{p+j}^p$ and the real constant $c$ determines the strip $a < \Ree \, s < b$ in the complex plane of  the Mellin variable $s$ where the
integrand in Eq.~(\ref{fin1}) is an analytical function. In our case,
$0 < c <1$. Explicitly, the Mellin momenta $\Omega_p(s)$ and
$R_j(s)$ in Eq.~(\ref{fin1}) read as
 \ba &&
\left(\frac{\alpha}{\pi}\right)^p\Omega_p(s)=
  \int_0^1 dx \; x^{2s} (1-x)^{1-s}
  \left[ \Pi^{(L)} \left( -\frac{x^2 }{1-x}m_L^2  \right)
  \right ]^p, \label{Omp1}
  \\[0.2cm] &&
\left(\frac{\alpha}{\pi}\right)^j R_j(s)= \int_0^\infty
\frac{dt}{t}\left( \frac{4m_{\ell}^2}{t}\right)^s \frac 1\pi \Imm
\left[ \Pi^{(\ell)} ( t)\right ]^j \label{rjj1}.
\ea
 The explicit expressions for  the operators
$\Pi^{(L)}$ and $\Pi^{(\ell)}$  in  Eqs.~(\ref{Omp1}) and (\ref{rjj1}) are
known in the literature, cf. Ref.~\cite{Lautrup:1969fr}:
\ba && \Ree \; \Pi^{(L,\ell)} (t) = \left(\frac{\alpha}{\pi}\right )
\left[\frac89 - \frac{\delta^2}{3}+\delta \left(\frac12
-\frac{\delta^2}{6}\right)\ln\frac{|1-\delta|}{1+\delta}\right],
\label{reP1} \\[0.2cm] &&
\frac1\pi \Imm  \;  \Pi^{(L,\ell)}(t) = \left(\frac{\alpha}{\pi}\right ) \delta
\left(\frac12 -\frac16 \delta^2\right)\theta\left (t-4m^2_{(L,\ell)}\right),
\label{imP1}
\ea
where    $\delta=\sqrt{1- 4m^2_{(L,\ell)}/{t}}$.
It is worth mentioning that due to the Euclidean character of the argument of $\Pi^{(L)}$ in Eq.~(\ref{Omp1})
and due to the $\theta$-function in Eq.~(\ref{imP1}),
the operators $\Pi^{(L)} \left( -\frac{x^2 }{1-x}m_L^2  \right)$  are pure real, do not depend on
the lepton mass and can be written as
 \ba &&
\Pi^{(L)} \left( -\frac{x^2 }{1-x}m_L^2  \right)=\frac\alpha\pi \left
[ \frac59 +\frac{4}{3x} - \frac{4}{3x^2}+ \left (-\frac13
+\frac{2}{x^2}-\frac{4}{3x^3}\right ) \ln(1-x)\right]. \label{lashk}
\ea
Furthermore, it is easy to show that
 $ R_j(s)$ is mass independent as well. Consequently, the only dependence on the lepton masses
enters  into $a_L$ in Eq.~(\ref{fin1})  through the ratio
 \ba
 r=\frac{m_\ell}{m_L}. \label{ratio}
 \ea
Accordingly, in the literature, it is commonly accepted to
classify the contributions to $a_L$ by this ratio (see, e.g., Ref.~\cite{review-2021}),
\begin{equation}\label{aL}
a_{L} =  ~A_{1}\left( \frac{m_L}{m_L}\right ) + ~A_{2}\left(
\frac{m_{\ell}}{m_L}\right )  +
A_{3}\left(\frac{m_{\ell_1}}{m_L},\frac{m_{\ell_2}}{m_L}\right ).
\end{equation}
The coefficients $A_1(r)$, being mass independent ($r=1$),    are universal for any type of lepton
and represent the set of diagrams with one, two, \ldots loops each formed by leptons $\ell$ of the
same type as the external one, $\ell=L$. It also includes the diagrams with no loops at all. Hence, the lowest order of the radiative corrections for $A_1(r)$ is $\alpha$. The coefficients
$A_2(r)$ stand for   diagrams with at least one loop  $\ell\neq L$ and, consequently, the lowest order of corrections
for them is $\alpha^2$. Analogously, the lowest order of corrections for $A_3(r_1,r_2)$ is $\alpha^3$ where $r_1= m_{\ell_1} / m_L,~r_2= m_{\ell_2}/m_L$, and  $m_{\ell_1}$ and $m_{\ell_2}$  are the masses of the two internal leptons $\ell_1\neq\ell_2\neq L$. Accordingly, one can write
 \ba
 && A_1(r=1)= A_1^{(2)}\left(\frac\alpha\pi
\right)^1 + A_1^{(4)}\left(\frac\alpha\pi \right)^2 +
A_1^{(6)}\left(\frac\alpha\pi \right)^3 + \cdots ,
\label{A1}\\[1mm]
&& A_{2}\left({r} \right) =
A_{2}^{(4)}(r)\left(\frac\alpha\pi \right)^2 +
A_{2}^{(6)}(r)\left(\frac\alpha\pi \right)^3 +
A_{2}^{(8)}(r)\left(\frac\alpha\pi \right)^4 +A_{2}^{(10)}(r)\left(\frac\alpha\pi
\right)^5 + \cdots,
\label{A23} ~~ \\[1mm]
&& A_3\left(r_1,r_2\right)= A_3^{(6)}(r_1,r_2)\left(\frac\alpha\pi
\right)^3 + A_3^{(8)}(r_1,r_2)\left(\frac\alpha\pi \right)^4 +
A_3^{(10)}(r_1,r_2)\left(\frac\alpha\pi \right)^5 + \cdots .\label{A3}~~~
\ea

In Eqs.~(\ref{A1})--(\ref{A3}),  the superscripts of $A_i^{(2n+2)}$ point to the corresponding order of the corrections from diagrams with the insertion of the vacuum polarisation operator with $n$ loops ($n=0,~1,~2,\ldots$), while the powers of $\alpha^{n+1}$  correspond to the  number $n$ of loops in the bubble-like Feynman diagrams. In what follows, for the coefficients $A_2^{(10)}(r)$ considered in this paper,  we adopt the notation $A_2^{(10) LL\ell\ell}(r)$, where ``$LL\ell\ell$'' \ reflects that the   four loops are formed by two leptons $L$ and two $\ell$.

Some comments are in line here.  The coefficient $A_1^{(2)}$, i.e. diagrams with no loops at all, determines the leading order correction to $a_L$ which, as already mentioned, has been computed by Schwinger~\cite{Schwinger1948} and found to be $a_L=\alpha/2\pi$. Hence, in Eq.~(\ref{A1}) the coefficient $A_1^{(2)}=\frac12$. The next order coefficients $A_1^{(2n+2)}$  are also known for a rather large $n$~\cite{Samuel-n-bubble,Jegerlehner:2017gek,Petermann:1957hs,Sommerfield:1957zz,Laporta:1996mq,Laporta:2018ngl}. It is important to note that for bubble-like diagrams the corrections decrease with increasing  $n$ up to $n=7$ loops and then increase for larger $n$. Moreover, for $n\gg 1$ these coefficients increase factorially, cf. Ref.~\cite{latrup77,LST-2022}.

 In what follows, we need an explicit expression for the  universal coefficient ${{A}}_{1}^{(10)}$ which is
\begin{equation} \label{A10iniv}
{{A}}_{1}^{(10)} =-\frac{3 689 383}{656 100}-\frac{21 928\;\pi^4}{1
403 325}-\frac{~128\;\zeta(3)}{675}+\frac{~64 \;\zeta(5)}{9}
               \simeq~4.7090571603 \times 10^{-4} \, .
\end{equation}

In the present paper, we focus on the derivation of analytical expressions for the  coefficients $A_{2}^{(10)}(r)$, i.e.,   diagrams with four  loops formed by two different types of leptons,   $\ell\neq L$ and two leptons $\ell = L$. The combinations $L\ell\ell\ell$, $LLL\ell$ and $\ell\ell\ell\ell$ will be presented elsewhere~\cite{tobepublished}.
Note also that the  universal coefficient ${{A}}_{1}^{(10)}$, for which $r=1$, can be obtained as the limit $\lim_{r\to 1}A_{2}^{(10)}(r)$ and
compared with Eq.~(\ref{A10iniv}). The case of diagrams with loops formed by two leptons $\ell_{1,2}$ different from $L$ is not considered.

\begin{figure}[h!]
\centering
\includegraphics[width=0.35\textwidth]{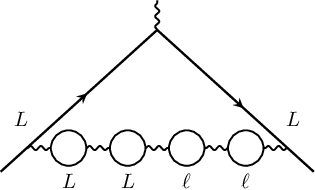}
\caption{The tenth order Feynman diagram contributing to the lepton $L$ anomalous magnetic momentum.
The two leptons $\ell$ are supposed to be different from $L$, $\ell\neq L$.
    }
   \label{4LLll}
\end{figure}

\section{Calculation of the diagram in Fig.~$\mbox{\boldmath\ref{4LLll}}$ }\label{glava2}

The Feynman diagram corresponding to the coefficients $A_2^{(10) LL\ell\ell}(r)$ is depicted in Fig.~\ref{4LLll}.

 For the diagram in Fig.~\ref{4LLll} one has: $p=2$  and  $j=2$ and, consequently, $F(2,2)=-6$. The Mellin momenta
 $\Omega_2(s)$ and  $R_2(s)$ are defined by Eqs.~(\ref{Omp1})--(\ref{imP1}). Calculations by parts of the integral
  (\ref{rjj1}) for $R_2(s)$ provide
 \ba R_2(s)=\frac{\sqrt{\pi}}{9} \; \frac{( s-1)(4s^2+13s+6)} {s^2
(s+2)(s+3)} \; \frac{\Gamma\left(1+s\right)}{\Gamma\left({\frac{3}{2}+s}
 \right)}\,.
 \label{R2}
\ea

As for the momentum $\Omega_2(s)$,   the presence of terms with $\ln^2(1-x)$  makes the integration  over $x$   much more involved.
To present the results in a more or less compact  form, we introduce several auxiliary functions containing integrals with powers of the logarithm $\ln^k (1-x)$, $k=0,~1,~2$:
\begin{equation}
X_k(s,n)=\int\limits_0^1  dx \; x^{2s+n} (1-x)^{1-s}\ln^k(1-x),\label{Xkk}
\end{equation}
Direct calculations of the integrals (\ref{Xkk}) result in, q.v.~Ref.~\cite{Solovtsova23}
\ba
&& X_0(s,n) =
\frac{\Gamma(2 - s) \Gamma(1 + n + 2 s)}{\Gamma(3 + n + s)} ,   \, X_1(s,n)=X_0(s,n)
\bigg( \psi \left( 2-s \right) -\psi \left( 3+n+s \right) \bigg),~~     \nonumber \label{XX}
 \\ [-0.2cm]
 &&
 \\
 &&
X_2(s,n)=
 X_0(s,n)
\left[   \bigg( \psi \left( 2-s \right)  -   \psi \left( 3+n+s \right)  \bigg) ^{2}+ \;
\psi^{(1)} \left(2-s \right) - \; \psi^{(1)} \left(3+n+s\right)  \right] . ~~
\nonumber
\ea
In Eqs.~(\ref{XX})   $\psi(s)$  and $\psi^{(1)}(s)$ are  the polygamma functions   of the zeroth and first
orders, respectively.
Then, the Mellin moment $\Omega_2(s)$ is explicitly expressed in terms of the introduced functions
(\ref{XX}) as
\ba
&& \Omega_2(s)\;  = \;
\frac{25}{81} X_0(s, 0) + \frac{16}{9}  X_0(s, -4) - \frac{32}{9} X_0(s, -3) +
 \frac{8}{27}  X_0(s, -2) + \frac{40}{27}  X_0(s, -1) -
  \nonumber \\ [0.1cm] &&
  \frac{10}{27} X_1(s, 0) +
 \frac{32}{9}  X_1(s, -5) - \frac{80}{9}  X_1(s, -4) + \frac{104}{27}  X_1(s, -3) +
\frac{ 28}{9}  X_1(s, -2) -  \frac{8}{9}  X_1(s, -1) +
 \nonumber \\ [0.1cm] &&
 \frac{1}{9} X_2(s, 0) +\frac{16}{9}  X_2(s, -6) -
 \frac{16}{3}  X_2(s, -5) + 4 X_2(s, -4) +
 \frac{8}{9}  X_2(s, -3) - \frac{4}{3}  X_2(s, -2).
\label{om2}
\ea
\vskip 0.2cm

Inserting $R_2(z)$ and  $\Omega_2(z) $ into Eq.~(\ref{fin1}), the final Mellin integral for $a_L$
acquires the form

\begin{equation}
a_{L}^{L L \ell  \ell}(r) = \left(\frac\alpha\pi\right)^5\frac{
-6}{2\pi i}\int\limits_{c-i\infty}^{c+i\infty} r^{-2s}{\cal F} (s)
ds\, , \label{fin4LL}
\end{equation}
where the integrand  ${\cal F} (s)$ is
\begin{eqnarray}
{\cal F} (s)& = & -\frac{8}{81} \pi^3\frac{(4s^2+13s+6)K_1(s) \cos(\pi
s)} {(s-2)\; Z_1(s) \sin^3(\pi s)} -  \frac{4}{27} \pi^2\left[ s \;
K_2(s) \left(-\psi^{(1)}({-s})+\frac {\pi^2}{2} \right) \right.
\nonumber \\[0.3cm]
 &&  \left.  + \frac
{1}{54}(1+2s)K_3(s)\right] \frac{(s-1)(4s^2+13s+6)}{\;Z_2(s) \sin^2(\pi s)} \, .
\label{F4LL}
\end{eqnarray}
For the sake of brevity the following notation has been introducded
\begin{eqnarray}
K_1(s)&=&24 + 74 s + 33 s^2 - 66 s^3 - 40 s^4 + 7 s^5 + 4
s^6,\nonumber
\label{Z1} \\
[0.2cm] K_2(s) &=& \left(2+3 s+s^2\right)^2 \left(-60-5 s+86 s^2-36
s^3+3 s^4\right),\nonumber
\label{Z2} \\
[0.2cm] K_3(s)&=& -864 - 39708 s - 77016 s^2 + 5072 s^3 + 69789 s^4 +
24717 s^5 -
 10761 s^6
\nonumber\\[0.1cm]
 &&
- 5025 s^7 + 384 s^8 + 256 s^9 ,\nonumber
\label{Z4}\\
[0.2cm]
Z_1(s)& =& s^3 (s+1)^2 (s+2)^3 (s+3)(2 s+1),\label{Y}  \\[0.2cm]
Z_2(s)&=&\;
(s+1)(s+2)(2s-1)(2s-3)(2s-5) Z_1(s). \label{YY}
\end{eqnarray}

It can be seen that the  integrand (\ref{F4LL}) is singular  at   integer values
of $s=\, \pm n$ and at some half-integer $s$, all singularities being  of the
pole-like nature. Consequently, the integral (\ref{fin4LL})
can be carried out by the Cauchy theorem closing the integration contour
in the left  ($r<1$) or   right ($r>1$) semiplanes of the Mellin complex variable
$s$ and computing the corresponding residues in these domains.

\subsection{Left semiplane,  $r< 1$}\label{subsec2}

In the left semiplane, the pole-like singularities  of the integrand
(\ref{F4LL}) arise owing to the functions $Z_1(s)$ and $Z_2(s)$,
Eqs.~(\ref{Y})--(\ref{YY}), and to the powers of  $\sin(\pi s)$ in the
denominators of Eq.~(\ref{F4LL}). For  integer $n<4$   the denominator
with $Z_1(s)\sin^3(\pi s)$ determines poles at $s=-n$ of the sixth
order for $n=0,2$,  of the fifth order for $n=1$, and poles of the
fourth order for $n=3$. Analogously, one counts the poles from the
second denominator with $Z_2(s)\sin^2(\pi s)$. Both terms in
Eq.~(\ref{F4LL}) have poles at $s=-n$ in the interval $(3< n\le
\infty)$ determined only by the corresponding powers of $\sin(\pi s)$.
Also, there is a single pole at $s=-\frac12$. The residues at
$s=0,-1,-2,-3$ and $s=-\frac12$ are calculated directly. As for
$s=-n$, $n>3$, the residues have been calculated in the most general
form by using a computer symbol manipulation package, e.g. {\it
Wolfram Mathematica}.
Next, applying Cauchy residue theorem, we obtain the following expression
\noindent
\begin{eqnarray}
&&  { {\ A}}_{2}^{(10),L L \ell \ell}(r<1) =
\frac{6874477399}{1052493750}-\frac{33715357}{1403325}\,r^2-
\frac{13586722367}{618866325}\,r^4-\frac{26021034733}{24960941775}\,r^6
   \quad \quad
\nonumber\\[0.2cm]
 &&
~~~ +\frac{256}{2025}\,r^6\ln(r)^3- \bigg[
\frac{47726}{6075}-\frac{4408}{2025}r^2+\frac{263306}{18225}r^4+\frac{80}{243}r^6
+\frac{8}{9}\left( \frac{4}{45}+\frac{130}{189} {r}^{4}-\frac
{96}{385}{r}^{6}\right) {\pi}^{2}
\nonumber \\
[0.2cm]
 &&
~~~~ - {\frac {64}{9}}\, \left( 1+{r}^{4}\right) \zeta(3)- {\frac
{8}{9}}\, \left( \frac{1}{3}+4{r}^{2}-\frac{13}{3}{r}^{4} \right) {\rm
Li_2}(r^2) - {\frac {256}{675}}\ln  \left( {\frac {1+r}{1-r}} \right)
r\bigg]\ln^2(r)
\nonumber \\
[0.2cm]
 &&
 ~~~~~ +  \bigg[
-{\frac {1131588179}{35083125}}-{\frac {3342554}{200475}}\,{r}^{2}+{
\frac {267396476}{9823275}}\,{r}^{4}-{\frac
{7215464}{7203735}}\,{r}^{6} + {\frac {8}{81}}\left( {\frac
{1810111}{92400}}+{\frac {48}{25}} r \right.
\nonumber \\
[0.2cm]
 &&
 \left.
 ~~~ - {\frac {3271}{154}}{r}^{2}+{\frac {3500711}{291060}}{r}^{4}-{ \frac
{1103962}{1334025}}{r}^{6} \right) {\pi }^{2} -{\frac {16}{81}} \,
\left(1+ {r}^{4} \right) {\pi }^{4} +{\frac {8}{81}}\, \left(
\frac{6}{5}+72\;{r}^{2}-{\frac {806}{7}}{r}^{4} \right.
 \nonumber \\
[0.2cm]
 &&
 \left.
 ~~~  +{ \frac {5184}{385}}{r}^{6} \right) \zeta(3)+{\frac {8 }{9}}\,
\left( {\frac {6733}{450}}+5\,{r}^{4}-{\frac {8}{75}}\,{r}^{
6}-\frac{54}{3}{r}^{2}+2\left(1+ {r}^{4}\right) {\pi }^{2} \right)
{\rm Li_2}(r^2)
  \bigg] \ln(r)
 \nonumber \\
[0.2cm] &&
 ~~~~ +   \bigg[{\frac
{148921}{30375}}+{\frac {176}{81}}\,{r}^{2}-{\frac {3727}{5103
}}\,{r}^{4}-{\frac {10856}{18711}}\,{r}^{6}+\frac{8}{3} \left({\frac
{1}{180}}+\frac{1}{3}\;{r}^{2}-{\frac {403}{756}}\,{r}^{4}+{\frac
{24}{ 385}}\,{r}^{6} \right) \pi^2 \nonumber
\\[0.2cm]
 &&
~~~~ - \frac{8}{3} \left( 1-2\,{r}^{2}+{r}^{4}-{\frac
{8}{225}}\,{r}^{6} \right) \ln(r) \bigg] {\rm Li_2}(1-r^2) -
 \left({\frac {512}{675}}\,\ln  \left( r \right) +{\frac {167}{540}}{\pi
}^{2}\right) r
 \bigg[ \left({\rm
Li_2}\frac{1-r}{1+r} \right)
 \nonumber \\
[0.2cm] &&
 ~~~~ -{\rm
Li_2}\left(-\frac{1-r}{1+r}\right)\bigg]+\bigg[ {\frac
{65786}{1002375{r}^{5}}}-\frac {460}{1701{r}^{3}}+\frac {58}{7r}
+\left( {\frac {127}{17820r^5}}-\frac {23}{756{r}^{3}}+\frac {29}{28r}
\right)\pi^2 \bigg]
 \nonumber \\
[0.2cm] &&
 ~~~~\times  \bigg[ {\rm
Li_2}\left(\frac{1-r}{1+r} \right)-{\rm
Li_2}\left(-\frac{1-r}{1+r}\right)- \frac{\pi ^2}{4}\bigg]
+\frac{1}{r^4} \bigg[ {\frac {131572}{1002375}}+{\frac {64}{675}}{\rm
Li_3}(r^2) + \left( -{\frac {131572}{1002375}} \right.
 \nonumber
\end{eqnarray}
\begin{eqnarray}
&&  ~~~~  \left. -{\frac {1339}{44550}}\,{\pi }^{2} -\frac{64}{675}
\left( {\rm Li_2}(r^2) -{\rm Li_2}(1-r^2)\right) \right)  \ln(r)
+\frac{127}{8910}\; \pi^2 \bigg] + {\frac {1024}{675}}\,r\;{\rm
Li_3}(-r)
 \nonumber \\
[0.2cm]
 &&  ~~~~ - \frac{1}{r^2}
\bigg[ {\frac {39221516}{63149625}}+{\frac { 16633}{280665}}\,{\pi
}^{2} + \frac {128}{675}\ln^2(r)-
 \left( {\frac {14455676}{21049875}}+{\frac {5248}{93555}}\,{\pi }^{2
} \right) \ln(r)  \bigg]
 \nonumber \\ [0.2cm]
 &&  ~~~~ -
\bigg[ \frac{26932}{2025} +\frac{128}{675} r - 16 r^2 +\frac {40}{9}
 r^4-{\frac{64}{675}}{r}^{6} + \frac{16}{9} \left(1+ {r}^{4}\right)
 \big(
{\pi }^{2}-2\;\ln^2 (r) \big) +{\frac {8}{9}}\, \left( {\frac
{14}{15}}
 \right. \nonumber \\
[0.2cm]
 &&
~~~~ \left. + 12{r}^{2}  - 13{r}^{4} \right) \ln(r)
 \bigg]{\rm Li_3}(r^2)+ \left[\frac{4}{5}+\frac{32}{3} r^2
-\frac{104}{9} r^4-\frac{32}{3} \left(1+\frac{4}{3} r^4\right)
\ln(r)\right]{\rm Li_4}(r^2)\nonumber \\
[0.2cm]
 &&
 ~~~~ +\frac{32}{3}
\left(1+\frac{5}{3}r^4\right) {\rm Li_5}(r^2) + \frac{8}{3} \left(
{\frac {4378771}{9355500}}+{\frac {4804663}{4365900}}\;{r}^{2}-{\frac
{3693612107}{4950930600}}\;{r}^{4} \right.
\nonumber \\
[0.2cm]
 &&
 ~~ \left.  + {\frac {49195554871}{499218835500}}\;
{r}^{6}\right)\pi^2 + \frac{8}{3}\left[{\frac {2683}{1350}}+{\frac
{166}{35}}\,{r}^{2}+{\frac {188033}{119070 }}\,{r}^{4}+{\frac
{1781036}{12006225}}\,{r}^{6} +\frac{4}{9} \left( 1+r^4 \right) \pi^2
 \right]
\nonumber \\
[0.2cm]
 &&
~~ \times \zeta(3)-\frac{8}{9} \left( {\frac {121}{5400}}-{\frac
{167}{1920}}\,r+{\frac {5}{18}}\,{r}^{2}-{ \frac
{1391}{4536}}\,{r}^{4}+{\frac {4}{1925}}{r}^{6} \right)
 \pi^4
+\frac{64}{9} (1+r^4) \zeta(5) + \Sigma_1(r),
\label{A10LL-A}
\nonumber \\
[-0.2cm]
 &&
\end{eqnarray}
where $\zeta$ is the Euler-Riemann zeta function, $\Phi$ is the Hurwitz--Lerch zeta function,
Li$_n$ are the polylogarithm functions of the order $n$  and $\Sigma_1(r)$
reflects the summation  over the remaining residues due solely to
$\sin^2(-\pi n)$  and $\sin^3(-\pi n)$ in Eq.~(\ref{F4LL}) for $n>3$;
\begin{eqnarray}
&& \Sigma_1(r) = \frac{8}{9}\;\sum_{n=4}^{\infty} \left[
\frac{B_1(-n)}{ Y(-n)}\; \psi_n^{(1)}+B_2(-n) \left(
2\psi_n^{(1)}\ln(r) + \psi_n^{(2)}\right) \right]
\frac{r^{2n}}{n(n-2)\;Y(-n)} \,, \label{SumLL-A}
\end{eqnarray}
where $\psi_n^{(1,2)}$ are the polygamma functions of  integer
arguments  $n$. Besides, in  Eq.~(\ref{SumLL-A}) the following
notation has been introduced:
\begin{eqnarray}
B_1 (n) &=& -64800-131220 n+521010
n^2+1261824n^3-516067n^4-2424104n^5 \nonumber \\
&&-9866n^6 +2083192n^7+23003n^8-842648n^9+41244n^{10} \nonumber \\
&& + 133968
n^{11}-9392 n^{12}-7104 n^{13}+576 n^{14} , \label{B1B2Y}
\\
B_2 (n) &=& (1-n)(6+13n+4n^2)(-60-5n+86n^2-36n^3+3n^4) \,,
\nonumber \\
Y(n) &=& -n(n+1)(n+2)(n+3)(2n+1)(2n-1)(2n-3)(2n-5)\,.
\nonumber
\end{eqnarray}

\subsection{Right semiplane, $r>1$}\label{subsec3}

In the right semiplane, the integrand  ${\cal F} (s)$ has pole-like singularities at half-integer $s$,  $s=\frac12,~\frac32,~\frac52$, and   an infinite number of poles at integer $s$, $s=1,~2,~3,~\cdots$. As in the previous case, we separate terms with poles exclusively  from $\sin(\pi s)$ and compute residues for  $s=\frac{1}{2},~\frac{3}{2},~\frac{5}{2},~1,~2$ directly. The remaining residues at integer $s=n$, $n>2$ are again calculated by means of the {\it Mathematica Wolfram} package. N.B.:   In the right semiplane, the polygamma function $\psi^{(1)}(-s)$ in Eq.~(\ref{F4LL})
has poles of the second order at integer $s=n$  as follows:
\ba
-\psi^{(1)}(-s)= \psi^{(1)}(s)-\frac{1}{s^2}-\frac{\pi^2}{\sin^2(\pi
s)},
\ea
where function $ \psi^{(1)}(s)$ is regular in the right semiplane.

Gathering all residues together,  we get
\begin{eqnarray}
&& \quad { {\ A}}_{2}^{(10), L L \ell  \ell}(r>1)
  =-\frac{229025509}{16839900}+\frac{1427197}{4009500 \;r^4}-\frac{16468}{200475
 } \;r^2-\frac{50510662\;
r^2}{1403325}+\frac{399848}{467775}\;r^4
\nonumber \\
 [0.2cm]
 &&  
~~~~+\pi^2 \left(-\frac{1801}{18711}-\frac{25291}{891000\;
r^4}+\frac{69763}{841995\;r^2}-\frac{98474}{93555}\; r^2+\frac{64
}{1155}\; r^4 \right) -\frac{136}{675\;r^4}\;
\zeta(3)
\nonumber \\
[0.2cm]
 &&
 ~~~~+\frac{16}{2025\; r^4} \ln^3(r) -\frac{8}{2025} \left( {\frac {492067}{308}}+{\frac {635}{44\; r^4}}-{\frac
{16637}{1386\;r^2} }+{\frac {288458}{77}}\,{r}^{2}-{\frac
{9264}{77}}\,{r}^{4} \right) \ln^2(r)
\nonumber \\
[0.2cm] && ~~~~-\frac{8}{2025} \left( {\frac {4314187}{924}}-\frac
{9463}{198\; r^4}-\frac {2533}{66\;r^2} +{\frac
{6333406}{693}}\,{r}^{2}-{\frac {99962}{231}}\,{r}^ {4} \right) \ln(r)
\nonumber \\
[0.2cm]
 &&
~~~~+ \frac{8}{81}\left( -{\frac {7433}{3696}}-{\frac
{3853}{13200\; r^4}}-{\frac {656}{ 3465 \; r^2}} -{\frac
{34729}{2310}}\,{r}^{2}+{\frac {432}{385}}\,{r }^{4}
 \right) \ln(r) \left( \pi^2 +4\ln^2(r) \right)
\nonumber \\
[0.2cm]
 &&
~~~~ -2 r \left( V_1(r) \left( \frac{\pi^2}{6} +2\ln^2(r) \right)+
V_2(r)\right) \bigg[ {\rm Li_2}\left(\frac{r-1}{r+1} \right)-{\rm
Li_2}\left(-\frac{r-1}{r+1}\right)- \frac{\pi ^2}{4}\bigg] -{8}{r}
\nonumber \\
[0.2cm] && ~~~~ \times \bigg(\frac{64}{675}-V_1(r) \ln(r) \bigg)
\bigg[ {\rm Li_3} \left(\frac{1}{r} \right)- {\rm
Li_3}\left(-\frac{1}{r}\right)\bigg] +{8}{r}\; V_1(r) \bigg[ {\rm
Li_4} \left(\frac{1}{r} \right)- {\rm
Li_4}\left(-\frac{1}{r}\right)\bigg]
\nonumber \\
[0.2cm]
 &&
~~~~ + ~\frac{8}{3}\left(\frac{32}{225}-V_1(r) \ln(r)\right)r
\ln^2(r)\bigg[
\ln\left(1+\frac{1}{r}\right)-\ln\left(1-\frac{1}{r}\right)\bigg]+V_3(r)\nonumber \\
[0.2cm]
 &&
~~~~\times\left \{ 2 \bigg[ {\rm Li_2}\left(\frac{r-1}{r+1} \right)-{\rm
Li_2}\left(-\frac{r-1}{r+1}\right)\bigg]- 4{\rm
Li_2}\left(1-\frac{1}{r}\right)
+\frac{\pi^2}{6}\right\}+\left(-8+\frac{64}{675 \;r^4}+16 r^2 \right.
\nonumber \\
[0.2cm]
 &&
 ~~~~ \left. -  \frac{32}{9 }r^4+\frac{64}{675 }r^6+\frac{16}{27}\pi^2 (r^4+1)
   \right)\bigg[ {\rm Li_2}\left(\frac{1}{r^2}\right)\ln(r)+{\rm
Li_3}\left(\frac{1}{r^2}\right)\bigg]+
\bigg[\left(-\frac{136}{405}-\frac{160}{27} r^2 \right.
\nonumber \\
[0.2cm]
 &&
 ~~~~\left.
+ \frac{13000}{1701} r^4-\frac{512}{1155}
r^6\right)\ln^2(r)+\frac{64}{27}(r^4+1)\ln^3(r)\bigg]{\rm
Li_2}\left(\frac{1}{r^2}
\right)+\bigg[\left(-\frac{128}{135}-\frac{128}{9}r^2\right.
\nonumber \\
[0.2cm]
 &&
 ~~~~\left.+ \frac{9776}{567}\;r^4-\frac{256}{385}\;r^6
\right) \ln(r)+\frac{32}{3}(r^4+1)\ln^2(r)\bigg]{\rm
Li_3}\left(\frac{1}{r^2} \right)+\bigg(
-\frac{124}{135}-\frac{112}{9}r^2+
\frac{8164}{567}r^4\nonumber \\
[0.2cm]
 &&
 ~~~~-\frac{128}{385}r^6+\frac{224}{9}(r^4+1)\ln(r)
\bigg){\rm Li_4}\left(\frac{1}{r^2}\right)+\frac{224}{9}(r^4+1) {\rm
Li_5}\left(\frac{1}{r^2}\right)\,.
 \label{A10LL-B}
\end{eqnarray}
where the term $\Sigma_2(r)$ originates from the summation of residues from $\sin(\pi s)$ at $s=n>2$;
\begin{eqnarray}
&& \Sigma_2(r)= \frac{8}{9}\;\sum_{n=3}^{\infty} \left[
\frac{B_1(n)}{ Y(n)}\; \psi_n^{(1)}+B_2(n) \left(
2\psi_n^{(1)}\ln(r) - \psi_n^{(2)}\right) \right]
\frac{r^{-2n}}{n(n+2)\;Y(n)} \,.
\label{SumLL-B}
\end{eqnarray}
Here $B_1(n)$, $B_2(n)$ and $Y(n)$ are defined in Eq.~(\ref{B1B2Y})
and the shorthand notation in Eq.~(\ref{A10LL-B}) is
 \ba  &&
 V_1(r)= {\frac {167}{540}}-{\frac {127}{17820}}\,{r}^{-6}+{\frac {23}{756}}\,{
r}^{-4}-{\frac {29}{28}}\,{r}^{-2}, \nonumber
 \\[0.2cm]  &&
 V_2(r)= {\frac {167}{135}}+{\frac
{254}{8019}}\,{r}^{-6}-{\frac {23}{189}}\,{r }^{-4}-{\frac
{256}{675}}\,\ln  \left( r \right), \nonumber
 \\[0.2cm]  &&
 V_3(r)= - {\frac {328}{243}}+{\frac
{32}{81}}\,{r}^{2}+{\frac {2168}{5103}}\,{r }^{4}-{\frac
{8248}{13365}}\,{r}^{6} +\pi^2 \left(-{\frac {2}{405}}-{\frac {8}{27}}\,{r}^{2}+{\frac
{806}{1701}}\,{r}^{4 }-{\frac {64}{1155}}\,{r}^{6}\right)\nonumber
 \\[0.2cm]  &&
~~~~~~~ - \left(\frac{8}{3}+\frac{64}{675 r^4}+16 r^2 - \frac{32}{9
}r^4+\frac{64}{675 }r^6\right)\ln  \left( r \right)~+4\left( -{\frac {2}{405}}-{\frac {8}{27}}\,{r}^{2}+{\frac
{806}{1701}}\,{r}^{4 } \right.
\nonumber
 \\[0.2cm]  &&
 \left.
~~~~~~~-{\frac {64}{1155}}\,{r}^{6}\right)\ln^2(r). \nonumber
 \ea

\section{Numerical results}\label{numerical}

The above Eqs.~(\ref{A10LL-A})--(\ref{SumLL-B}) represent the exact analytical
expressions of the tenth order of the radiative corrections from diagrams with the
insertion of four lepton loops, as depicted in Fig.~\ref{4LLll}. Despite
their cumbersomeness, the explicit analytical  form  allows for numerical
calculations with any desired precision. The precision can only be limited by the knowledge of the basic
physical constants $\alpha$, $m_\ell$ and $m_L$.

\begin{figure}[ht]
\includegraphics[width=0.6\textwidth]{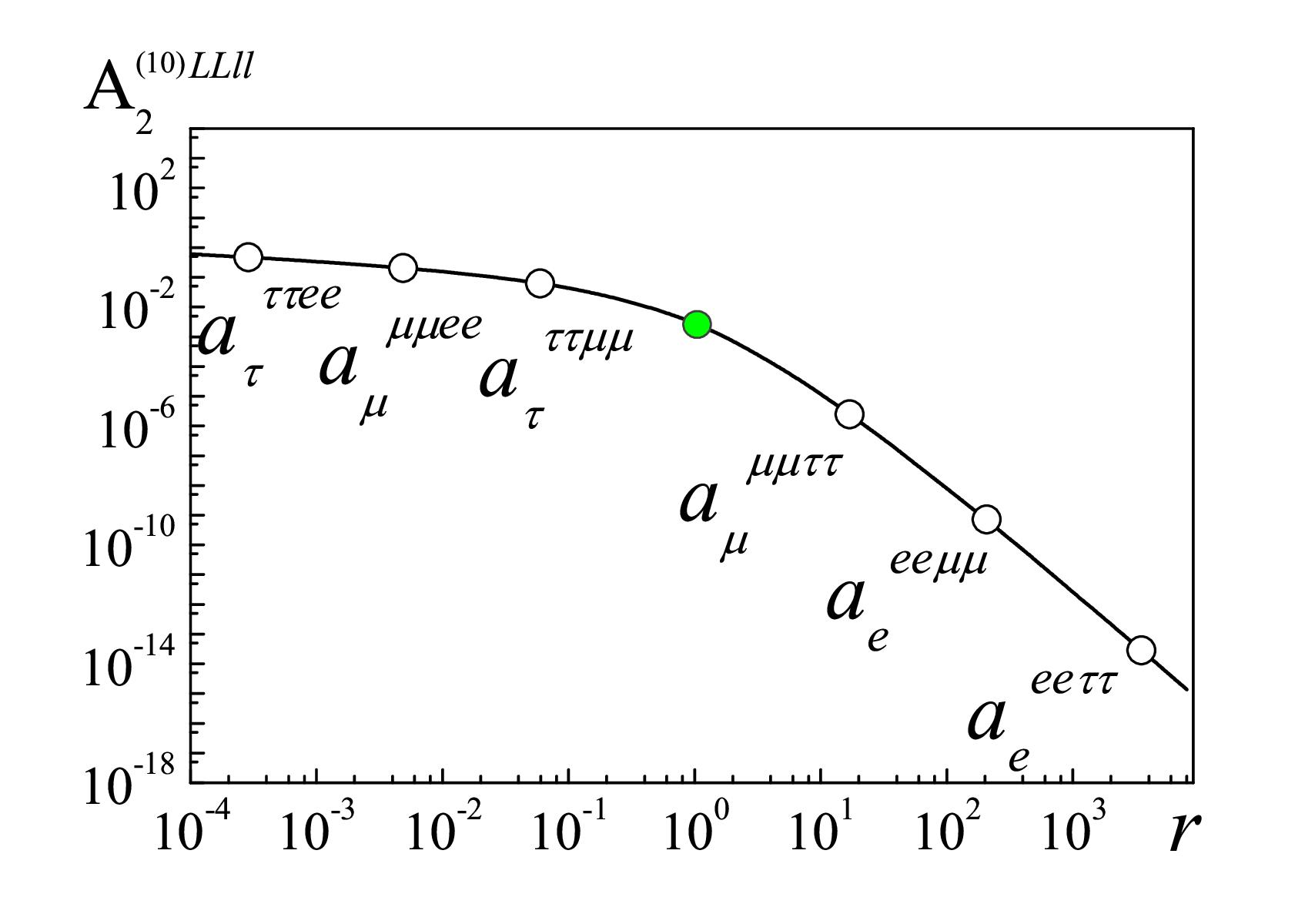}\hspace*{1cm}
\caption{The tenth order coefficients $A_2^{(10) LL\ell\ell}(r)$,
Eqs.~(\ref{A10LL-A})--(\ref{SumLL-B}), as a function  of the ratio $r
= m_\ell/m_L$, for the external lepton $L$ with the insertions of the
polarizationoperators with four loops. The notation $A_2^{(10)
LL\ell\ell}(r)$  indicates that the four loops correspond to two
leptons  $L$ and two leptons  $\ell$ ($L,\ell = e, \mu$ and $\tau$).
The open circles, as well as the associated with them labels, point to
physical values of the ratio $r$ and to the corresponding physical
coefficients $A^{(10) LL\ell\ell}(r)$. The universal value
$A_2^{(10)}(r=1)$  is displayed by the full circle.
\label{A410Numeric}}
\end{figure}

The results of numerical calculations are presented in Fig.~\ref{A410Numeric} for the coefficients ${ {\ A}}_{2}^{(10), L L \ell  \ell}(r)$, Eq.~(\ref{A23}), which are displayed as a function of $r$ (solid line), where  $r$  ranges  in the interval $0 <  \, r\, <\, \infty$. The open circles, accompanied by the corresponding labels $a_{L}^{LL\ell\ell}$, denote all the possible combinations of the real existing leptons $L$ and $\ell$ ($L,\ell = e,\mu$ and $\tau$). The full circle corresponds to $r=1$, i.e., to the universal coefficient $  A_{2}^{(10), L L \ell \ell}(r=1)=6A_1^{(10)}\simeq 0.00282543429618$, cf. Eq.~(\ref{A10iniv}).

It can be seen from Fig.~\ref{A410Numeric} that the corrections $A_{2}^{(10), L L \ell \ell}(r)$ for the anomaly of the lepton $L$ basically come from the region $r<1$, i.e. from diagrams with loops formed by leptons $\ell$ lighter than the external one. Moreover, they are larger than the universal coefficients at $r=1$. Contributions from loops with leptons $\ell$ heavier than the lepton $L$ rapidly decreasing  with increase  $r$ and can be by 10--11 orders of magnitude below the universal value. Note that, since the large scale of the ordinate axis, more than 18 orders of magnitude), Fig.~\ref{A410Numeric} illustrates only qualitatively the main features of the tenth order corrections. For precise calculations of $A_{2}^{(10), L L \ell \ell}(r)$, one should use directly the analytical forms
presented in Eqs.~(\ref{A10LL-A})--(\ref{SumLL-B}).

\section{Asymptotic expansions}\label{asym}

As mentioned, the tenth order corrections in the literature have been considered analytically only as limiting expansions   $r \ll 1$~cf. Refs.~\cite{Aguilar:2008qj,Laporta94} and only for muons, i.e., only diagrams with electron internal loops  have been included into the consideration. Since  Eqs.~(\ref{A10LL-A})--(\ref{A10LL-B}) provide exact analytical expressions for the QED corrections from the diagrams in Fig.~\ref{4LLll}, we can calculate the corresponding  limits not only at  $r \ll 1$ but also at $r \gg 1$ for any type of leptons $L$ and $\ell$. The results of expansions can  be then compared with the expansions  presented in Refs.~\cite{Aguilar:2008qj,Laporta94}. To make the comparison with Refs.~\cite{Aguilar:2008qj,Laporta94} easier, below we present the asymptotic expansions in terms of the variable $t\equiv r^2$. The corresponding expansions of Eqs.~(\ref{A10LL-A})--(\ref{SumLL-B}) read as:

\noindent

\vskip 1.0cm

 \paragraph{{\rm Eq.~(\ref{A10LL-A}),  $t=r^2 \ll 1$.}}
\begin{eqnarray}
&&   A_{2,asymp.}^{(10)}(t) \underset{t \ll 1} =
\;\left(-\frac{943}{486}-\frac{8}{405}\pi^2+\frac{16}{9}\zeta(3)\right)\ln^2
t-\left(\frac{57899}{7290}-\frac{10766}{6075}\pi^2+\frac{8}{81}\pi^4
-\frac{8}{135}\zeta(3)\right)
\nonumber \\
&& ~~~ \times \ln t
-\frac{1090561}{109350}-\frac{148921}{91125}\pi^2-\frac{106}{6075}\pi^4
+\frac{10732}{2025}\zeta(3)+\frac{32}{27}\pi^2\zeta(3)+\frac{64}{9}\zeta(5)
\label{asympR<1t}  \nonumber \\ [0.2cm]
&&+\left[-\left(\frac{1832}{243}-\frac{104}{315}\pi^2-\frac{32}{9}\zeta(3)\right)\ln
t
-\frac{619798}{25515}+\frac{564008}{297675}\pi^2-\frac{8}{81}\pi^4+\frac{1328}{105}\zeta(3)\right]t
\nonumber\\
&&+\left[\left(\frac{29696}{45927}+\frac{852346}{535815}\pi^2-\frac{8}{81}\pi^4-\frac{3224}{567}\zeta(3)\right)\ln
t-\left(\frac{470}{729}+\frac{260}{1701}\pi^2-\frac{16}{9}\zeta(3)\right)\ln^2
t  \right.\nonumber \\
&& +
\left.\frac{51445307}{28934010}-\frac{546693856}{168781725}\pi^2+\frac{26}{729}
\pi^4+\frac{752123}{178605}\zeta(3)+\frac{32}{27}\pi^2\zeta(3)+\frac{64}{9}\zeta(5)\right]t^2
\nonumber\\ [0.2cm]
&&+ \left[ \frac{244420473203}{72937816875} - \frac{186003297484
\pi{^2}}{374414126625} + \frac{64 \pi^4}{2475} -
\left(\frac{94530074}{105249375}-\frac{7124144\pi^2}{108056025} -
\frac{256}{385}
 \zeta(3) \right) \right.
 \nonumber \\
  &&
\left. \times \ln(t) - \left(\frac{11114}{ 30375} -
   \frac{64 \pi^2}{1155} \right)\ln^2(t ) + \frac{32}{2025}\ln^3(t)
 + \frac
   {14248288}{36018675}\zeta(3)\right]  t^3 + {\cal
   {O}}\left(t^4\ln^4 t \right).
 \label{asympR<1}
\end{eqnarray}

\vspace*{3mm} Our expansion (\ref{asympR<1}) can be directly compared
term by term with their analogues presented in
Refs.~\cite{Aguilar:2008qj,Laporta94}.  A meticulous inspection of
expansions (\ref{asympR<1t}) with their analogues in
Ref.~\cite{Aguilar:2008qj} (Eq.~(A13)) and Ref.~\cite{Laporta94}
(Eq.~(3)) shows that the term $\frac{64}{9}\zeta(5)$ is absent in
Ref.~\cite{Aguilar:2008qj}. Besides,  there is another misprint in
Ref.~\cite{Aguilar:2008qj}, namely the term

 \vspace*{2mm}
 \noindent
$$-\left[\left(\frac{470}{729}+\frac{260}{1701}\pi^2-\frac{16}{9}\zeta(3)\right)\ln^2 t \right]t^2
$$
has been omitted.

\begin{figure}[hbt]
\includegraphics[width=0.49\textwidth]{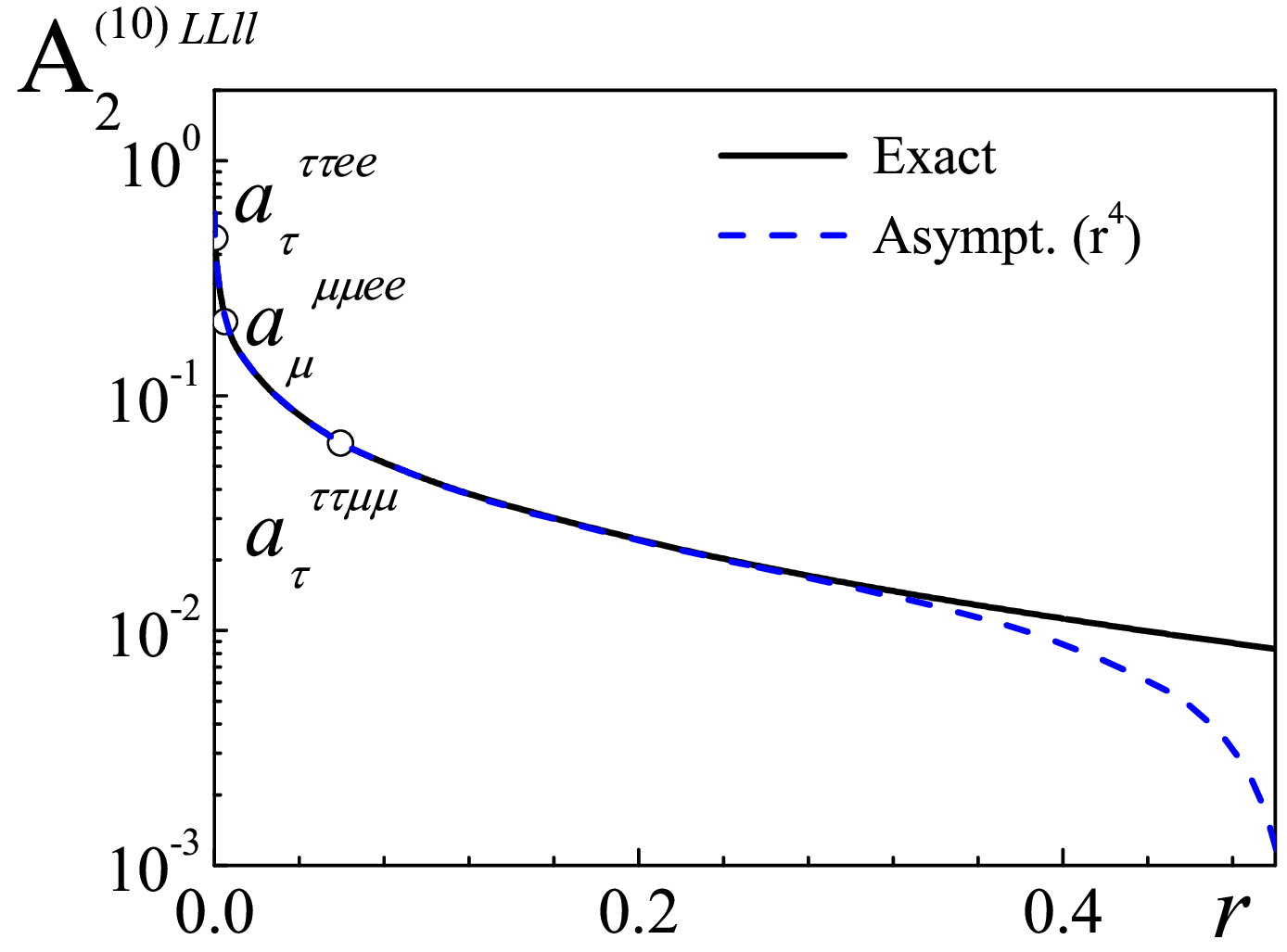} 
\includegraphics[width=0.475\textwidth]{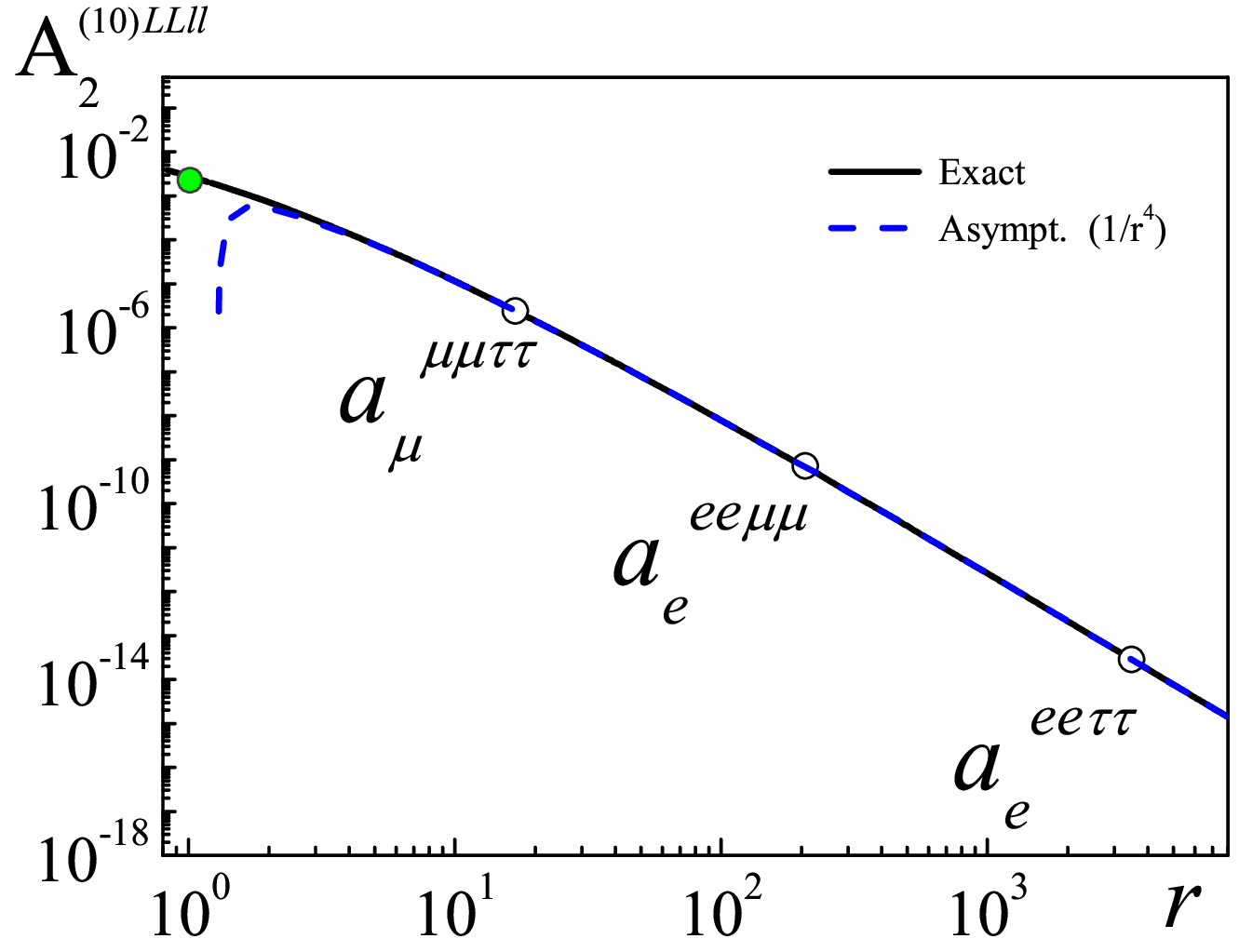}
\caption{The asymptotic expansion of the coefficients
$A_2^{(10)\ LL\ell\ell} $,  dashed lines, vs. the exact calculations by
Eqs.~(\ref{A10LL-A})--(\ref{SumLL-B}), solid lines.
Left panel:  $r<1$; Right panel  $r>1$. The meaning of the open circles and labels
$a_L^{LL\ell\ell}$ is the same as in Fig.~\ref{A410Numeric}.
} \label{RisAsymp}
\end{figure}

 \paragraph{{\rm Eq.~(\ref{A10LL-B}),  $t=r^2\gg 1$.}}

 \vspace*{5mm}
 \noindent
\begin{eqnarray}
&&
 A_{2,asymp.}^{(10)}(t) \underset{t \gg 1} = \;\left(\frac{6938293}{29160000}
 +\frac{20327}{2430000}\ln t - \frac{61}{81000}\ln^2 t+\frac{2}{2025}\ln^3
t-\frac{136}{675}\zeta(3)\right)\frac{1}{t^2}  \label{asympR>1} \\[0.3cm]
&&
+\left(\frac{1364352509}{87516450000}+\frac{675182}{52093125}\ln t -
\frac{6073}{1488375}\ln^2 t+\frac{2}{1375}\ln^3 t
-\frac{8}{525}\zeta(3)\right)\frac{1}{t^3} 
  + {\cal
{O}}\left(\frac{1}{t^4}\right) \nonumber .
\end{eqnarray}

\vspace*{5mm}

The comparisons of the limiting  expansions  $r\to 0$, Eq.~(\ref{asympR<1t}),
and $r\to\infty$, Eq.~(\ref{asympR>1}), with exact numerical
calculations by Eqs.~(\ref{A10LL-A})--(\ref{SumLL-B}) are presented in Fig.~\ref{RisAsymp}.
One can conclude that the approximate expansions  practically coincide with the exact formulae
in quite large intervals of $r$, namely  $(0<r<0.2)$ for the expansion~(\ref{asympR<1})
and $(4<r< \infty)$ for the expansion (\ref{asympR>1}), herewith both intervals include
all the corresponding physical values of $a_L^{LL\ell\ell}$. It means that in these
intervals qualitative estimates of the corrections to any type of leptons can be safely
performed by asymptotic expansions  instead of the exact but cumbersome expressions Eqs.~(\ref{A10LL-A})--(\ref{SumLL-B}).

\section{Summary}\label{summ}

In summary,  we have derived, for the first time,  analytical
expressions of  the QED corrections $A_2^{(10)}(r)$ of the tenth order
to the anomalous magnetic moment of leptons $L$ ($L = e , \mu$  or
$\tau$ ) generated by  diagrams with insertions of four loops in the
photon vacuum polarization operator. We considered the diagram with
loops formed by  two leptons $L$ of the same type as the external one
and two leptons $\ell$ different from $L$. The radiative corrections
have been obtained  in close analytical forms in terms of the mass
ratio $r=\frac{m_\ell}{m_L}$.   The generic variable $r $ is defined
in the whole region $(0 < r < \infty)$. Our consideration is based on
a combined use of the dispersion relations for the vacuum polarization
operators and the Mellin--Barnes integral transform for the Feynman
parametric integrals. This  technique is widely used in the literature
in multi-loop  calculations in relativistic quantum field theories,
c.f.~Refs.~\cite{Friot:2005cu,Rafael-HVP,Kotikov:2018wxe}. The final
integrations have been performed by the Cauchy residue theorem in the
left ($r<1$) and right ($r>1$) semiplanes of the complex Mellin
variable $s$. We investigated numerically the behaviour of the
corrections $A_2^{(10)}(r)$ in the whole interval  $0 < r < \infty$
for all possible combinations of the two leptons $L$ and two $\ell$.
We found that the contributions to $A_2^{(10)}(r)$  basically come
from the region $r<1$, i.e. from diagrams with loops formed by leptons
$\ell$ lighter than the external one. Moreover, these corrections  are
larger than the universal coefficients at $r=1$. Contributions from
loops with leptons $\ell$ heavier than the lepton $L$ rapidly decrease
with increasing  $r$ and can be by 10--11 orders of magnitude below
the universal value.

We have compared our analytical expressions the
corresponding asymptotic  expansions with the well-known results available
in the literature and found that our results are fully compatible with the
earlies known calculations. We also argued that the asymptotic expansions in
the regions $r<0.2$ and $r>4$ are rather close to the exact values and can
serve as reliable estimates of $A_2^{(10)}(r)$ in these regions.

\section*{Acknowledgments}
This work was supported in part by a grant
of the JINR--Belarus collaborative program.

\newpage



\end{document}